\documentclass[12pt]{article}
\usepackage{amsxtra,amssymb,amsthm,amsmath,latexsym}

\theoremstyle{plain}
\newtheorem{theorem}{Theorem}

\newtheorem*{question}{Question (Problem) P}
\newtheorem*{problem1}{Problem P$_1$}
\newtheorem{lemma}{Lemma}

\newcommand{\refT}[1]{Theorem~\ref{T:#1}}
\newcommand{\refS}[1]{Section~\ref{S:#1}}

\newcommand{\refL}[1]{Lemma~\ref{L:#1}}

\def\R{{\mathbb R}}

\def\oH{\buildrel\circ\over H}
\def\oH1{\buildrel\circ\over H\kern-.02in{}^1}

\def\l{\ell}
\def\barh{\overline{h}}

\def\bysame{\rule{.5in}{.005in},\ }

\def\const{\hbox{\,const\,}}

\def\ve{{\varepsilon}}

\def\be{\begin{equation}}
\def\ee{\end{equation}}  
\def\bee{\begin{equation*}}
\def\eee{\end{equation*}}

\begin{document}


\title{ Completeness of the set of scattering amplitudes
   \thanks{Key words: completeness, scattering amplitude, 
            approximation, inverse problems, nanotechnology, smart 
materials}
   \thanks{AMS subject classification: 
         35J10, 35P25, 35R30, 74J25, 81V05}
   \thanks{PACS: 0230, 0340K, 0365 }
}

\author{
A.G. Ramm\\
Mathematics Department, 
Kansas State University, \\
 Manhattan, KS 66506-2602, USA\\
ramm@math.ksu.edu\\
}

\date{}

\maketitle\thispagestyle{empty}

\begin{abstract} 
Let $f\in L^2(S^2)$ be an arbitrary fixed function on the unit sphere
$S^2$, with a sufficiently small norm, and $D\subset \R^3$ be an arbitrary 
fixed bounded domain. 
Let $k>0$ and $\alpha\in S^2$ be fixed.

It is proved that there exists a potential $q\in L^2(D)$ such that 
the corresponding scattering amplitude
$A(\alpha')=A_q(\alpha')=A_q(\alpha',\alpha,k)$ approximates
$f(\alpha')$ with arbitrary high accuracy:
$\|f(\alpha')-A_q(\alpha')_{L^2(S^2)}\|\leq\ve$,
where $\ve>0$ is an arbitrarily small fixed number.
The results can be used for constructing
nanotechnologically "smart materials".
\end{abstract}


\section{Introduction}
Let $D\subset\R^3$ be a bounded domain, $k=\const$, $\alpha\in S^2$,
$S^2$ is the unit sphere. Consider the scattering problem:
\be\label{e1} [\nabla^2+k^2-q(x)]u=0\hbox{\quad in \quad }\R^3, \ee
\be\label{e2}
  u=u_0+A_q(\alpha',\alpha,k)\frac{e^{ikr}}{r}+o\left(\frac{1}{r}\right),
  \quad u_0=e^{ik\alpha\cdot x}, \quad r=|x|\to\infty,
  \quad \alpha'=\frac{x}{r}, \ee
The coefficient $A_q$ is called the scattering amplitude, and $q\in 
L^2(D)$ is a potential. The solution to \eqref{e1}-\eqref{e2} is called 
the scattering solution. It solves the equation
\be\label{e3}
  \begin{aligned}
  u &=u_0-Tu, \qquad Tu:=\int_D g(x,y) q(y) u(y)dy,\\
    &\ g=g(x,y,k)=\frac{e^{ik|x-y|}}{4\pi|x-y|},
     \ \quad u=u(y)=u(y,\alpha,k). 
  \end{aligned}\ee
The solution $u$ to \eqref{e3} is unique in $L^2(D)$ for any 
complex-valued $q$ for which $\|T\|<1$, i.e., for any sufficiently small 
$q$.

We are interested in the following problem, which differs from the 
standard inverse scattering problem with fixed-energy data, studied in 
\cite{R}.

\begin{question}
Given an arbitrary fixed $f(\alpha')\in L^2(S^2)$,
can one find a $q\in 
L^2(D)$, such that 
\be\label{e4}
  \|A_q(\alpha')-f(\alpha')\|_{L^2(S^2)} \leq\ve,\qquad
  A_q(\alpha')=A_q(\alpha',\alpha,k),\ee
where $\ve>0$ is an arbitrary small number, $\alpha\in S^2$ and $k>0$ are 
fixed?
\end{question}

The answer to this question was not known. The scattering problem in 
\eqref{e1}--\eqref{e2} has been studied much (see, e.g., \cite{C}, 
\cite{P}, \cite{R}). The inverse scattering problem with fixed-energy data 
(ISP) was solved in \cite{R1}, \cite{R2}, \cite{R3}, and 
\cite[Chapter 5]{R}. This problem consists of finding $q\in L^2(D)$ from 
the corresponding scattering amplitude $A_q(\alpha',\alpha,k)$ given for 
all $\alpha',\alpha\in S^2$ at a fixed $k>0$. It was proved in \cite{R1} 
that this problem has a unique solution. In \cite{R3} reconstruction 
algorithms were proposed for finding $q$ from exact and from noisy data, 
and stability estimates were established for the solution.

The problem P we have posed in this paper is different from the ISP. The 
data $A_q(\alpha')$ do not determine $q\in L^2(D)$ uniquely, in general. 
The potential $q$ for which \eqref{e4} holds is not unique even if 
$\ve=0$ and $f(\alpha')=A_q(\alpha',\alpha)$ for some $\alpha\in S^2$ and 
$q\in L^2(D)$. We want to know if there is a  $q\in L^2(D)$, $q=q_\ve(x)$, 
such that \eqref{e4} holds with an arbitrarily small fixed $\ve>0$.

We prove that the answer is yes, provided that $f$ is sufficiently small. 
The "smallness" condition will be specified in our proof. The question 
itself is motivated by the 
problem $P_1$ 
studied in \cite{R4}:

\begin{problem1}
Can one distribute small acoustically soft particles in a bounded domain 
so that the resulting domain would have a desired radiation pattern, i.e., 
a desired scattering amplitude?
\end{problem1}

The problem P, which is studied here, is of independent interest. 

The following two lemmas allow us to give a positive answer to Question P
under the "smallness" assumption.

\begin{lemma}\label{L:1}
Let $f\in L^2(S^2)$ be arbitrary and $k>0$ be fixed. Then
\be\label{e5}
  \inf_{h\in L^2(D)} \|f(\alpha')+\frac 1 {4\pi}\int_D 
e^{-ik\alpha'\cdot 
x} h(x)dx\|_{L^2(S^2)}= 0. \ee
\end{lemma}

\begin{lemma}\label{L:2}
Let $h\in L^2(D)$ be arbitrary, with a sufficiently small norm. Then
\be\label{e6}
  \inf_{q\in L^2(D)} \|h-qu\|_{L^2(D)}=0, \ee
where $u=u(x,\alpha,k)$ is the scattering solution corresponding to $q$.
Under the above "smallness" assumption, there exists a potential
$q$, such that $qu=h$.
\end{lemma}

If Lemmas 1 and 2 are proved, then the positive answer to Question P 
(under the "smallness" assumption) follows from the well-known formula for 
the scattering 
amplitude:
\be\label{e7}
  A_q(\alpha')=-\frac{1}{4\pi}\int_D e^{-ik\alpha'\cdot x}
  q(x)u(x,\alpha,k)dx, \ee
in which $k>0$ and $\alpha\in S^2$ are fixed. The answer is given in 
\refT{1}.

\begin{theorem}\label{T:1}
Let $\ve>0$, $k>0$, $\alpha\in S^2$ and $f\in L^2(S^2)$ be arbitrary, 
fixed, with a sufficiently small norm. Then there is a $q\in L^2(D)$ such 
that \eqref{e4} holds.
\end{theorem}

There are many potentials $q$ for which (4) holds.

In \refS{2} proofs are given.

In \refS{3} a method is given for finding $q\in L^2(D)$  such that 
\eqref{e4} holds.

\section{Proofs}\label{S:2}
\begin{proof}[Proof of \refL{1}]
If \eqref{e5} is not true, then there is an $f\in L^2(S^2)$ such that
\be\label{e8}
  0=\int_{S^2} d\beta f(\beta) \int_D e^{-ik\beta\cdot x}h(x)dx
   =\int_D dxh(x) \int_{S^2} f(\beta) e^{-ik\beta\cdot x}d\beta
   \quad \forall h\in L^2(D). \ee
Since $h(x)$ is arbitrary, relation \eqref{e8} implies
\be\label{e9}
  \int_{S^2} f(\beta) e^{-ik\beta\cdot x} d\beta=0
  \qquad  \forall x\in D. \ee
The left-hand side of \eqref{e9} is the Fourier transform of a compactly 
supported distribution $\frac{\delta(\lambda-k)}{k^2}f(\beta)$, where
$\delta(\lambda-k)$ is the delta-function. 
Since the Fourier transform is injective, it follows that $f(\beta)=0$.
This proves Lemma 1.

An alternative proof can be given. It is known that
\be\label{e10}
  e^{-ik\beta\cdot x} 
  = \sum^\infty_{\l=0, -\l\leq m \leq \l} 4\pi i^\l j_\l(kr) 
\overline{Y_{\l,m}(-x^0)} Y_{\l, m}(\beta),
  \qquad r:=|x|,\ x^0:=\frac{x}{r}, \ee
where $Y_{\l, m}$ are orthonormal in $L^2(S^2)$ spherical 
harmonics, $Y_{\l, m}(-x^0)=(-1)^{\l}Y_{\l, m}(x^0)$,
$ j_\l(r):=(\frac {\pi}{2r})^{1/2}J_{\l+\frac 1 2}(r)$, and $J_{\l}(r)$
is the Bessel function.
Let
\be\label{e11}   f_{\l, m}:=(f,Y_{\l, m})_{L^2(S^2)}. \ee
From \eqref{e9} and \eqref{e10} it follows that
\be\label{e12}   f_{\l, m} j_\l(kr)=0, \qquad \forall x\in D, -\l\leq 
m\leq \l. \ee
If $k>0$ is fixed, one can always find $r=|x|$, $x\in D$, such that 
$j_\l(kr)\not= 0$.

Thus, \eqref{e12} implies $f_{\l, m}=0\quad\forall \l, -\l\leq m\leq \l.$ 
\refL{1} is proved. The "smallness" assumption is not needed in this 
proof.
\end{proof}

\begin{proof}[Proof of \refL{2}]
In this proof we use the "smallness" assumption. If the norm
of $f$ is sufficiently small, the the norm of $h$ is small so that 
condition (23) (see below) is satisfied. If this condition is satisfied, 
then formula (24) (see below) yields the desired potential $q$,
and $h=qu$, where $u$ is the scattering solution, corresponding to $q$.
Therefore, the infimum in (6) is attained. Lemma 2 is proved.
\end{proof}

Let us give another argument, which shows the role of the  "smallness" 
assumption from 
a different point of view. Note, that if $||q||\to 0$, then the 
set of 
the functions $qu$ is a linear set. In this case, if one 
assumes that \eqref{e6} is not true, then one can claim that there is an 
$h\in L^2(D)$, $h\not= 0$, such that
\be\label{e13}  \int_D dx h(x)q(x)u(x)=0 \qquad \forall q\in L^2(D),
\quad ||q||<<1.\ee
Choose
\be\label{e14} q=c\barh e^{-ik\alpha\cdot x}, \qquad c=\const>0. \ee

Let $c$ be so small that $\|T\|=O(c)<1$, where 
 $T:L^2(D)\to L^2(D)$ is defined in \eqref{e3}. Then equation
\eqref{e3} is uniquely solvable and 

\be\label{e15}
   u=u_0+O(c)\qquad\hbox{\ as\ }c\to 0. 
\ee
From \eqref{e13} and \eqref{e15} one gets
\be\label{e16}
   c\int_D |h|^2 dx+O(c^2)=0 \qquad \forall c\in(0,c_0). \ee
If $c\to 0$, then \eqref{e16} implies $\int_D |h(x)|^2 dx=0$. Therefore $h=0$.

\section{A method for finding $q$ for which \eqref{e4} holds}\label{S:3}

Lemmas 1 and 2 show a method for finding a $q\in L^2(D)$ such that 
\eqref{e4} holds. Given $f(\alpha')\in L^2(S^2)$, let us find $h(x)$ such 
that
\be\label{e17}
  \|f(\alpha')+\frac 1 {4\pi}\int_De^{-ik\alpha'\cdot 
x}h(x)dx~\|^2_{L^2(S^2)} 
  <\ve^2.\ee
This is possible by \refL{1}. It can be done numerically by taking 
$h=h_n=\sum^n_{j=1} c_j \varphi_j(x)$, where 
$\{\varphi_j\}_{1\leq j<\infty}$ is a basis of $L^2(D)$, and minimizing 
the quadratic form on the left-hand side of \eqref{e17} with respect to 
$c_j,\ 1\leq j<n$. For sufficiently large $n$ the minimum will be less 
than $\ve^2$ by \refL{1}.

The minimization  with the accuracy $\ve^2$ can be done analytically: let 
$B_b$ be a ball of radius 
$b$ centered 
at a point $0\in D$, $B_b\subset D$, $h=0$ in $D\setminus B_b$, 
$$h(x)=\sum^\infty_{\l=0, -\l \leq m \leq \l} h_{\l, m}(r) Y_{\l, m}(x^0) 
\quad \hbox { in } B_b,$$ 
$$f(\alpha')=\sum^\infty_{\l=0, -\l \leq m \leq \l} f_{\l, m} 
Y_{\l, m}(\alpha'),$$ 
and 
$$\sum_{l>L, -\l \leq m \leq \l}|f_{\l, m}|^2<\ve^2.$$ 
For $0\leq \l \leq L$ let us equate the Fourier coefficients of
$f(\alpha')$ and of the integral from (17). 
We use the known formula:
$$e^{-ik\beta \cdot x}=\sum_{\l=0, -\l \leq m \leq \l} 4\pi 
i^{\l}j_{\l}(kr)\overline{Y_{\l, m}(-x^0)}Y_{\l, m}(\beta).$$
This leads to the 
following relations for finding $h_{\l, m}(r)$:
\be\label{e18}  f_{\l, m}=-(-i)^\l\sqrt{\frac {\pi}{2k}} \int^b_0 
r^{3/2} J_{\l+\frac 1 2}(kr)h_{\l, m}(r)dr, 
  \qquad 0\leq \l\leq L.  \ee
There are many $h_{\l, m}(r)$ satisfying equation (18). Let us take 
$b=1$ and use \cite[Formula 8.5.5]{B},
\bee\begin{aligned}
  \int^1_0 
  & x^{\mu+\frac 1 2} J_\nu(kx) dx\\
  & =k^{-\mu-\frac 3 2}
  \left[
(\gamma+\mu-\frac{1}{2})kJ_\nu(r)S_{\mu-\frac 
1 2, \nu-1}(k)-kJ_{\nu-1}(k)S_{\mu+\frac{1}{2}, \nu}(k) 
  +2^{\mu+\frac{1}{2}}
  \frac{\Gamma\left(\frac{\mu+\nu}{2}+\frac{3}{4}\right)}
       {\Gamma\left(\frac{\nu-\mu}{2}+\frac{1}{4}\right)}
  \right]\\
  &:=g_{\mu, \nu}(k), \end{aligned}\eee
where $S_{\mu,\nu}(k)$ are Lommel's functions. Thus, one may take 
$$h_{\l, m}(r)=\frac {f_{\l, m}} {-(-i)^{\l}\sqrt{\frac {\pi}{2k}}g_{1, 
\l+\frac 1 2}(k)}, \quad 0\leq 
\l\leq L;\quad h_{\l, m}(r)=0 
\quad \l>L.$$
Thus, the coefficients $h_{\l, m}(r)$ do not depend on $r\in (0,1)$
in this choice of $h$: inside the ball $B_b$ the function  
$h(x)$=$h(x^0)$ depends only on the angular variables, and outside this 
ball $h=0$.

With this choice of $h_{\l, m}(r)$ the left-hand side of \eqref{e17} 
equals to 
$$\sum^\infty_{\l=L+1, -\l \leq m \leq \l}|f_\l|^2<\ve^2.$$ 
The above choice of 
$h_{\l, m}(r)$, which yields an  analytical choice of $h(x)$, is one of 
the choices for which \eqref{e17} holds. The function $g_{1, \l+\frac 1 
2}(k)$ in the definition of $h_{\l, m}(r)$ decays rapidly when $\l$
grows. This makes it difficult numerically to calculate 
accurately  $h_{\l, m}(r)$ when $\l$ is large. If $f=1$ in a small solid 
angle and  $f=0$ outside of this angle, then one needs large $L$ to
approximate $f$ by formula \eqref{e17} with small $\ve$. Numerical
difficulties arise in this case. This phenomenon is similar to the one
known in optics and antenna synthesis as superresolution difficulties
(see \cite{R6},  \cite{R7}, \cite{R8}).

Let $h\in L^2(D)$ be a function for which \eqref{e17} holds. Consider the 
equation
\be\label{e19}  h=qu, \ee
where $u=u(x; q)$ is the scattering solution, i.e., the solution to 
equation \eqref{e3}. Let
\be\label{e20}
  w:=e^{-ik\alpha\cdot x}u,  \quad G(x,y):=g(x,y)e^{-ik\alpha\cdot(x-y)},
  \quad H:=he^{-ik\alpha\cdot x}, \quad \psi:=qw. \ee
Then equation \eqref{e3} is equivalent to the equation
\be\label{e21} w=1-\int_D G(x,y)\psi(y)dy, \ee
and $\psi(y)=H(y)$ by \eqref{e19}. Multiply \eqref{e21} by $q$ and get 
$$\psi (x)=q(x)-q(x) \int_D G(x,y)\psi(y)dy.$$ 
Since $\psi(x)=H(x)$, we 
get
\be\label{e22}
  q(x)=H(x) \left[1-\int_D G(x,y)H(y)dy\right]^{-1}. \ee
If $H(x)$ is such that
\be\label{e23} 
  \inf_{x\in D} \bigg| 1-\int_DG(x,y)H(y)dy\bigg|>0, \ee
then $q(x)$, defined in \eqref{e22}, belongs to $L^2(D)$ and \eqref{e4} 
holds. If $H$ is sufficiently small then \eqref{e23} holds. 
If $D$ is small and $H$ is fixed, then \eqref{e23} holds because
\bee
  \bigg|\int_D G(x,y)Hdy\bigg|
  \leq \int_D \frac{|H|dy}{4\pi |x-y|} \leq \frac{1}{4\pi} \|H\|_{L^2(D)} 
  \left(\int_D \frac{dy}{|x-y|^2}\right)^{1/2} =O(a^{1/2})\|H\|, \eee
where $a$ is the radius of a smallest ball containing $D$.

Formula \eqref{e22} can be written as
\be\label{e24}
  q(x)=h(x) \left[u_0-\int_D g(x,y)h(y)dy\right]^{-1}, \ee
where $u_0$ and $g$ are defined in \eqref{e3}.
Numerically equation \eqref{e24} worked for $f(\beta)$ which were large 
in absolute value.

\section{An idea of a method for making a material with the desired
radiation pattern}

Here we describe an idea of a method for calculation of a 
distribution of small particles,
embedded in a medium, so that the resulting medium would have a desired
radiation pattern for the plane wave scattering by this medium.
This idea is described in more detail in [R4]. The results of this paper
complement the results in [R4], but are completely independent of [R4],
and are of independent interest.

Suppose that a bounded domain $D$ is
filled in by a homogeneous material. Assume that we embed many small
particles into $D$. Smallness means that $k_0a<<1$, where $a$ is the 
characteristic 
dimension of a particle and $k_0$ is the wavenumber in 
the region $D$ before the small particles were embedded in $D$.
The question is: 

{\it Can the density of the distribution of these particles in $D$  be
chosen so that the resulting medium would have the desired radiation 
pattern
for scattering of a plane wave by this medium?}

For example, if the direction $\alpha$ of the incident
plane wave is fixed, and the wave number 
$k$ of the incident plane wave $e^{ik\alpha \cdot x}$ 
in the free space outside $D$ is fixed, then
can the particles be distributed in $D$ with such a density
that the scattering amplitude $A(\alpha', \alpha, k)$
of the resulting medium would approximate with the desired accuracy an 
a priori given arbitrary function $f(\alpha')\in L^2(S^2)$ in the sense 
(4)?
 
Assume that the particles are acoustically soft,
that is, the Dirichlet boundary condition is assumed on their boundary, 
that $k_0a<<1$ and $\frac a d<<1$, where $d>0$
is the minimal distance between two distinct particles,
and that the number $N$ of 
the small particles tends to infinity in such a way that
the limit  $C(x):= \lim_{N\to \infty}\lim_{r\to 0}\frac {\sum_{D_j\subset 
B(x,r)} C_j}{|B(x,r)|}$ exists.
Under these assumptions, it is proved in [R4] that the scattered field can 
be described 
by  equation (1) with $q(x)$  that can be expressed analytically via
$C(x)$. Here $D_j$ is the region occupied by $j-$th acoustically 
soft particle, $C_j$ is the 
electrical capacitance of the perfect conductor with the shape $D_j$,
$B(x,r)$ is the ball  of radius $r$ centered at the point $x$,
and $|B(x,r)|=\frac {4\pi r^3} 3$ is the volume of this ball.   
If the small particles are identical, and $\mathcal{C}$ is the 
electrical capacitance of one small paticle, then 
$C(x)=\mathcal{N}(x)\mathcal{C}$, where  $\mathcal{N}(x)$ is the number
of small particles per unit volume around the point $x$. Formulas for the 
electrical capacitances for conductors of arbitrary shapes are derived in 
[R5].


\begin{thebibliography}{x}

\bibitem[B]{B} H. Bateman, A. Erdelyi, Tables of integral transforms,
McGraw-Hill, New York, 1954.

\bibitem[C]{C}
H.~Cycon, R.~Froese, W.~Kirsch, B.~Simon,
{\it Schr\"odinger Operators}, Springer, Berlin, 1986.

\bibitem[P]{P}
D.~Pearson, 
{\it Spectral Theory}, Acad. Press, London, 1988.

\bibitem[R]{R}
A.~G.~Ramm,
{\it Inverse Problems}, Springer, New York, 2005.

\bibitem[R1]{R1} 
\bysame
 Recovery of the potential from fixed energy
 scattering data. Inverse Problems, 4, (1988),
 877-886; 5, (1989) 255.

\bibitem[R2]{R2} 
\bysame
 Stability estimates in inverse scattering,
Acta Appl. Math., 28, N1, (1992), 1-42.


\bibitem[R3]{R3} 
\bysame
 Stability of solutions to inverse scattering
problems with fixed-energy data, Milan Journ of Math.,
70, (2002), 97-161.

\bibitem[R4]{R4} 
\bysame
 Distribution of particles which produces a
desired radiation pattern,
Communic. in Nonlinear Sci. and Numer. Simulation,
(to appear).

\bibitem[R5]{R5}
\bysame 
{\it Wave scattering by small bodies of
arbitrary shapes}, World Sci. Publishers, Singapore, 2005.

\bibitem[R6]{R6}
\bysame
Apodization theory I, II, III,  Optics and Spectroscopy, 27, (1969),
508-514; 29, (1970), 390-394; 29, (1970), 594-599.

\bibitem[R7]{R7}
\bysame
 On resolution ability of optical systems,  Optics and Spectroscopy,
29, (1970), 794-798.

\bibitem[R8]{R8}
\bysame
Optimal solution of antenna synthesis problem, Doklady Acad. of Sci.
USSR, 180, (1968), 1071-1074.




\end{thebibliography}
\end{document}